\documentclass[twocolumn, prl, amssymb,  aps, preprintnumbers,
amsmath,floatfix]{revtex4}

\usepackage{epstopdf}
\usepackage{graphics}
\usepackage{graphicx}
\usepackage{dcolumn}
\usepackage{bm}
\usepackage{longtable}
\usepackage{epsfig}
\usepackage{times}
\usepackage{url}
\usepackage{color}

\begin{document}
\title{Magnetic Quantum-Phase Control between Two Entangled Macroscopic Nuclear Ensembles}

\author{Wen-Te \surname{Liao}}
\email{Liao@mpi-hd.mpg.de}

\author{Christoph H. \surname{Keitel}}

\author{Adriana \surname{P\'alffy}}
\email{Palffy@mpi-hd.mpg.de}

\affiliation{Max-Planck-Institut f\"ur Kernphysik, Saupfercheckweg 1, D-69117 Heidelberg, Germany}
\date{\today}
\begin{abstract}
Heralded generation and manipulation of quantum entanglement between two macroscopic and spatially separated crystals at room temperature is theoretically studied. 
We show that by combining an x-ray parametric down-conversion source and x-ray interferometry with nuclear resonant scattering techniques, two macroscopic crystals hosting M\"ossbauer nuclei located each on an interferometer arm can be entangled for few tens of nanoseconds. The coherence time of the entanglement state can be prolonged up to values comparable to the lifetime of a single nuclear excited state, on the order of hundred nanoseconds.
A non-mechanical magnetic control of the quantum phase between the two spatially separated entangled  nuclear crystals is put forward.
\end{abstract}
\maketitle

The ability to create entanglement between quantum memories in a heralded manner \cite{Usmani2012} is vital for quantum communication developments,  in particular for quantum repeaters \cite{Duan2001} and quantum networks \cite{Kimble2008}.
The search for scalable quantum repeaters has come up with solid-state resources, which require entanglement between quantum memories hosted in spatially separated macroscopical crystals \cite{Usmani2012,Tittel2010}. So far, such macroscopical entanglement has been typically limited in either time duration, working temperature or sample size/number of atoms involved, as shown in Table~\ref{table1} that lists some key achievements. As generic feature, these experiments  make use of optical photons as entanglement carriers \cite{Julsgaard2001,Chou2005,Lee2011,Usmani2012}. Recent developments in x-ray optics \cite{PfeifferWGuide,Shvydko2003,Shvydko2004,Chang2005,JarreWGuide,Chen2008,Shvydko2010,Shvydko2011,Osaka2013} and single x-ray quanta manipulation \cite{Liao2012a,Liao2014,Vagizov2014} have pointed out the advantages of higher frequency photons: deeper penetration, better focusing, robustness and improved detection \cite{Vagizov2014}, all of them pertinent for quantum technology applications. 

\begin{table}[b]
\vspace{-0.4cm}
\caption{\label{table1} Experimental parameters for demonstrated
entanglement between macroscopic objects. The case of the $^{57}$FeBO$_3$ crystal is under theoretical investigation in this work.
}
\center{
\begin{tabular}{lccccc}
\hline
Target                        & Temperature            & Coherence           & Distance   & Ref.                      \\
                              & (K)                    & time                &            &                           \\ 
\hline \\
Nd$^{3+}$Y$_2$SiO$_5$ crystal & 3                      &7      ns            &  1.3 $cm$  & \cite{Usmani2012}         \\
$10^{12}$ Caesium atoms       & 300                    &0.5    ms            &  few $cm$  & \cite{Julsgaard2001}      \\ 
$10^{5}$  Caesium atoms       & $< 1$                  &1      $\mu$s        &  2.8 $m$   & \cite{Chou2005}           \\ 
Diamond crystal               & 300                    &7      ps            &  15  $cm$  & \cite{Lee2011}            \\
$^{57}$FeBO$_3$ crystal       & 300                    & $\lesssim141$  ns          &  $\lesssim 1 m$   & this work       \\
\hline
\end{tabular}
}
\end{table}

The commissioning of the first x-ray free electron lasers (XFEL) \cite{slac,Sacla} and achievements in non-linear x-ray optics such as x-ray parametric down conversion (XPDC) \cite{Tamasaku2007,Tamasaku2011,Shwartz2011,Shwartz2012},
sum-frequency generation of x-ray and optical wave \cite{Glover2012}, and x-ray two-photon absorption \cite{Tamasaku2014} extend the frontiers of quantum optics towards higher frequencies. The field of x-ray quantum optics \cite{adams2013} 
promises exciting applications in metrology \cite{Schibli2008,Cavaletto2013,Cavaletto2014} and information technology \cite{Liao2012a,Liao2014b,Vagizov2014}, as well as for generation of quantum entanglement in the keV regime using nuclear rather than atomic transitions \cite{Palffy2009,Liao2014}. X-ray photons from nuclear transitions would also be ideal for the exploration of quantum correlations and entanglement of macroscopic bodies, although this direction was so far never addressed. 
In this Letter we put forward for the first time a scheme to create and manipulate heralded entanglement between two macroscopic solid objects, i.e., crystals containing M\"ossbauer nuclei, using x-rays at room temperature. We show that a setup comprising nuclear forward scattering (NFS) \cite{Hastings1991,Roehlsberger2004}, XPDC \cite{Tamasaku2007,Tamasaku2011} and x-ray interferometry \cite{Hasegawa1995,Hasegawa1999,Tamasaku2002} can be superior to  previously employed schemes providing 
longer coherence times ($\sim$100 ns), room temperature handling and larger samples ($\sim10^{18}$  atoms). Furthermore, an alternative function of the same setup involving lattice mechanical excitations in the crystal (phonons) opens the possibility to explore the boundary between quantum realm and the classical world \cite{Haroche1998,Zurek2003} and test decoherence models \cite{decoh}.

\begin{figure*}[ht]
\begin{center}
\vspace{-0.4cm}
  \includegraphics[width=1\textwidth]{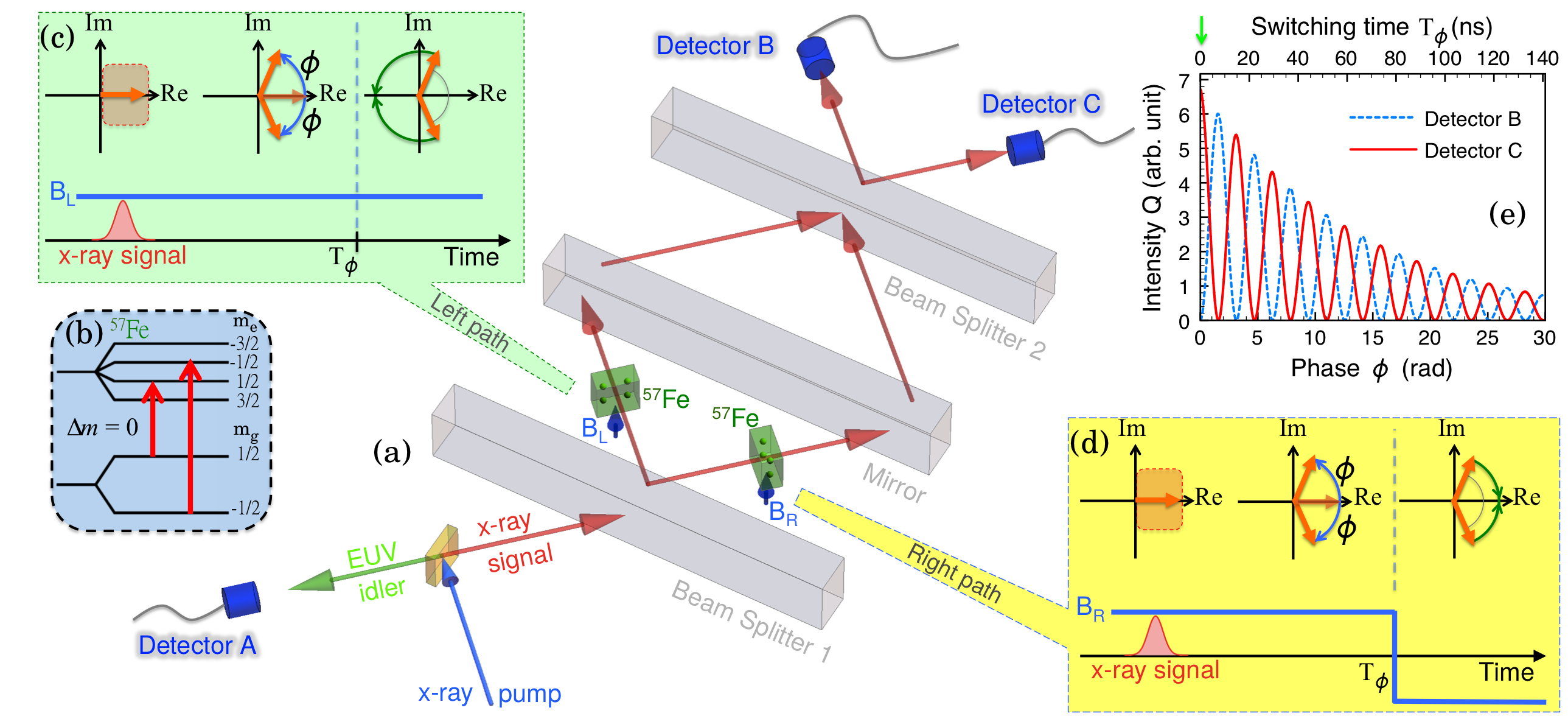}
  \caption{\label{fig1}
(Color online) (a) A combination of x-ray interferometry with nuclear resonant scattering and an XPDC setup. X $\rightarrow$ X + EUV down-conversion occurs within a diamond crystal (yellow cuboid). Subsequently, a converted single x-ray signal photon (red arrow) enters an x-ray interferometer while a converted EUV idler photon (green arrow) reaches detector A producing a click. Beam splitter BS 1 transfers the signal photon into a two-path entanglement state $\vert \mathrm{TPE}\rangle=(|1\rangle_{L}|0\rangle_{R}+ i |0\rangle_{L}|1\rangle_{R})/\sqrt{2}$. The $\vert \mathrm{TPE}\rangle$ photon is then
subject to NFS as it impinges on two $^{57}$Fe crystals (green slabs). The
nuclear transitions in the latter experience hyperfine splitting under the action of the applied magnetic fields $\bf{B_L}$ and $\bf{B_R}$ (blue short arrows).
As the $\vert \mathrm{TPE}\rangle$ single photon is absorbed and shared by the two distant nuclear crystals, the latter are entangled in the state $\vert \mathrm{ME}\rangle=(|E\rangle_{L}|G\rangle_{R}+ i |G\rangle_{L}|E\rangle_{R})/\sqrt{2}$.
The re-emitted signal photon from the nuclear crystals is in turn reflected by the mirror, recombined at beam splitter BS 2 and registered by either detector B or C. (b) $^{57}$Fe nuclear level structure. A linear polarized x-ray signal photon drives two $\Delta m=0$ transitions (red arrows). (c)(d) Dynamics of the nuclear currents (rotating orange arrows)
on left and right arm of the interferometer induced by the time-dependent magnetic fields $\bf{B_L}$ and $\bf{B_R}$ (blue solid lines), respectively. $\bf{B_R}$  is inverted at switching time $\mathrm{T_\phi}$. (e) Interference pattern Q (see text) at detector B and C for different $\mathrm{T_\phi}$. The light green downward arrow indicates the moment a click at detector A starts the chronometer for $\mathrm{T_\phi}$.
  }
\end{center}
\end{figure*}

In NFS experiments \cite{Hastings1991,Roehlsberger2004} 
a monochromatic x-ray pulse resonant to a nuclear transition energy, i.e.,  14.4 keV for $^{57}$Fe, coherently propagates through the M\"ossbauer solid-state sample and is detected in the forward direction. The NFS technique predates the XFEL  originally making use of synchrotron radiation (SR) pulses which are monochromatized to meV bandwidths. When impinging on a nuclear crystal, each SR pulse creates at most a single delocalized excitation known also as a nuclear exciton state 
$\vert \mathrm{E} \rangle=\frac{1}{\sqrt{N}}\sum_{\ell=1}^{N}e^{i\vec{k}\cdot \vec{r}_{\ell}}\vert g\rangle\vert e_{\ell}\rangle$.
Here,  the $\ell$th nucleus at position $\vec{r}_{\ell}$ is excited by the incident x-ray with the wave number $\vec{k}$, whereas  all other $(N-1)$ nuclei stay in their ground state \cite{Haas1988,Roehlsberger2004,Smirnov2005,Smirnov2007,Liao2014}. 
Intuitively, a single nuclear sample crystal can be divided into two remote parties labeled by $L$ and $R$, leading to the formation of the following entangled state $\vert \mathrm{ME} \rangle$  between two distant parties
\begin{eqnarray}
\vert \mathrm{ME} \rangle 
&=&\frac{1}{\sqrt{2}}(|\mathrm{E}\rangle_{L}|\mathrm{G}\rangle_{R}+ e^{i\phi} |\mathrm{G}\rangle_{L}|\mathrm{E}\rangle_{R})\, ,
\label{eq2}
\end{eqnarray}
where $|\mathrm{E}\rangle_{L(R)}$ and $|\mathrm{G}\rangle_{L(R)}$ stand for the $L$($R$) ensemble being in the excited state $\vert \mathrm{E} \rangle$ and in the ground state, respectively, and $\phi$ is the relative phase between the two components.
This corresponds to the single-photon version of quantum entanglement, i.e., the entanglement of two field modes by a single photon \cite{Enk2005}, that has been successfully used to create heralded quantum entanglement between two crystals with optical photons \cite{Usmani2012}.

The nuclear exciton in NFS has  remarkable properties that support the study of  quantum entanglement between macrosocopic samples: 
(1) the size of the nuclear sample crystals subject to x-ray irradiation is macroscopic with a typical dimension of mm$\times $mm$\times\mu$m \cite{Shvydko1996,Smirnov1996NE}.
(2) NFS experiments are typically performed at room temperature or even higher  temperatures reached in the sample due to the bombardment of x-rays. At room temperature, the formed nuclear exciton exhibits many quantum effects for a duration commensurate with the nuclear excited state lifetime, e.g., 141 ns for $^{57}$Fe \cite{Shvydko1996,Smirnov1996NE,Roehlsberger2004,Heeg2013}. 
(3) the collective dynamics of the exciton, e.g., its decay speed-up \cite{Hannon1999} and the coherent re-emission of single x-ray quanta with $\vec{k}$ directionality \cite{Roehlsberger2004}, can be manipulated by external fields \cite{Shvydko1996,Palffy2009,Liao2012a,Liao2012b,adams2013}.
(4) the efficiency of an x-ray detector for the considered energy is nearly 100\%, and the noise is very low \cite{Vagizov2014}.

Yet, a number of obstacles for entanglement detection arise in the traditional NFS setup. For instance, the coherent decay of the exciton in the forward direction requires the absence of spin flip and nuclear recoil making difficult to check whether excitation has occurred in the sample. The typical monitoring of the nuclear exciton is the detection of a time-delayed single x-ray photon that follows the decay of state $|\mathrm{E}\rangle$  \cite{Hasegawa1995,Hasegawa1999,Shvydko1996,Smirnov1996NE,Roehlsberger2004}, however with many instances without any signal.  Finally, for entanglement control and detection one may not be able to rely on the Raman technique as employed by the original DLCZ protocol \cite{Duan2001}, since in nuclear $\Lambda$-type systems the required initial nuclear state  is very challenging to prepare \cite{Liao2011,Liao2013}.

Properties (1-4) suggest that the NFS setup can provide a robust and controllable entanglement state $|\mathrm{ME}\rangle$ between two or more macroscopic nuclear crystals at room temperature.
To overcome the outlined disadvantages and prepare with certainty the entangled state $|\mathrm{ME}\rangle$, we adopt the heralded method \cite{Scully2006b,Usmani2012,Bruno2013} and put forward the scheme illustrated in Fig.~\ref{fig1}. An XPDC single-photon source provides a heralded x-ray photon resonant to the nuclear transition. In Fig.~\ref{fig1}(a), we use an x-ray $\rightarrow$ x-ray + extreme ultraviolet (X $\rightarrow$ X + EUV) \cite{Tamasaku2007,Tamasaku2011} setup consisting of a diamond crystal that splits the pump x-ray into an EUV idler and  an x-ray signal photon. 
Alternatively, an X $\rightarrow$ X + X down-conversion \cite{Shwartz2011,Shwartz2012} setup can also be used. The heralded time sequence of NFS is triggered by registering an EUV idler photon at the detector A. At the same time the x-ray signal photon enters an x-ray interferometer, e.g., triple Laue interferometer \cite{Hasegawa1995,Hasegawa1999}. After propagating through the 50/50 beam splitter BS 1 \cite{Osaka2013}, a two path entanglement state $\vert \mathrm{TPE}\rangle=(|1\rangle_{L}|0\rangle_{R}+ i |0\rangle_{L}|1\rangle_{R})/\sqrt{2}$ of the x-ray signal photon is formed and then interacts with two distant $^{57}$Fe crystals which are under the action of the hyperfine magnetic fields $\bf{B_L}$ and $\bf{B_R}$. Here, state $|1\rangle_{L(R)}$ and $|0\rangle_{L(R)}$ refer to the one-photon Fock state and the vacuum state at the left (right) path, respectively. Later on, the two  NFS paths are recombined on a 50/50 beam splitter BS 2 and monitored by detectors B and C. 
The key for the setup is the arrangement of x-ray and EUV detectors such that without interacting with the nuclear sample crystals, either both detectors A and B, or both detectors A and C simultaneously register the two XPDC photons. Each successful creation of the entanglement state $|\mathrm{ME}\rangle$ is heralded by the click at detector A while no photon is registered at detectors B, C (registering coherent decay of nuclear exciton) or any other detectors monitoring the $4\pi$ emission angle (for photon loss or incoherent, spontaneous decay of the nuclear exciton).  
The missing count of an x-ray signal photon is  attributed to the absorption by the two remote crystals.
The absorbed and rescattered signal photon reaches the detectors B or C only later, with a time delay on the order of the nuclear excited state lifetime, i.e., approx. 100 ns. Due to detection efficiency (though in principle very high for x-rays \cite{Vagizov2014}) or incoherent decay processes in $4\pi$ solid angle, the original lack of signal at 
detectors B and C might just be the consequence of photon loss. However, in this case the click at detector A will not be followed by any further photons registered with detectors B and C (the 14.4 keV photon background is negligible), such that no misleading events may be recorded.

The typical bandwidth  of the down-converted photons is around 1 eV~\cite{Tamasaku2007} corresponding to a pulse duration of 1 fs. This bandwidth is much broader than the linewidth of the interacting nuclear transition such that NFS can be treated as
the coherent propagation of an ultrashort pulse through the resonant nuclear media \cite{Crisp1970,Van1999}. Due to the hyperfine magnetic field, each $^{57}$Fe 14.4 keV nuclear transition is split into a sextet as illustrated in Fig.~\ref{fig1}(b). Linearly polarized x-rays will drive simultaneously the two $\Delta m=0$ transitions,  with Zeeman energy shifts $\pm\hbar\Delta_{\bf B}$. 
The  coherently scattered photon wavepacket off the nuclear crystals can be written as \cite{Crisp1970,Van1999,Hannon1999}
\begin{equation}
\psi(t)=\frac{\alpha}{\sqrt{\alpha\Gamma t}}J_{1}\left( 2\sqrt{\alpha\Gamma t}\right)\cos(\Delta_{\bf B}t)e^{-\frac{\Gamma}{2}t}\, . 
\end{equation}
Here, $J_1$ is the Bessel function of the first kind and of the first order \cite{Crisp1970,Van1999,Smirnov2005,Shvydko1999N} caused by multiple scattering \cite{Shvydko1999N} or dispersion \cite{Crisp1970,Van1999}, $\alpha$ the effective resonant thickness \cite{Shvydko1999N} and $\Gamma$ the spontaneous decay rate of the nuclear excited state. Furthermore, the trigonometric oscillation is caused by the quantum beat of the two split nuclear transitions \cite{Shvydko1998,Van1999}, and the exponential decay term describes the incoherent spontaneous decay of the excited states. 
The coherently re-emitted x-ray signal photon will then be reflected by an x-ray mirror with near 100\% reflectivity \cite{Shvydko2011} and subsequently recombined on  beam splitter BS 2. Finally, either detector B or C will register 
the time-delayed x-ray signal from the two output ports of the interferometer with probabilities that depend on the relative phase between the two paths.

To verify the entanglement between the two nuclear sample crystals, we invoke the method of quantum state tomography \cite{Chou2005,Laurat2007,Usmani2012} to determine the density matrix $\widetilde{\rho}$ of $\vert \mathrm{TPE}\rangle$ of the coherently re-emitted single x-ray photon from two targets.  In the photon-number basis , $\widetilde{\rho}$ reads \cite{Chou2005,Laurat2007}
\begin{eqnarray}\label{dm}
\widetilde{\rho} &=& \frac{1}{P}
\left( 
\begin{array}{cccc}
 p_{00}  & 0       & 0         & 0      \\
 0       & p_{01}  & d_{tpe}   & 0      \\
 0       & d_{tpe}^* & p_{10}  & 0      \\
 0       & 0       & 0         & p_{11} 
\end{array}  
\right),
\end{eqnarray}
where $p_{ij}$ is the probability of detecting $i$ photons  from the left crystal and $j$ photons  from the right one. Furthermore, $d_{tpe}$ is the coherence between the  two components of $\vert \mathrm{TPE}\rangle$ and $P=\mathrm{Tr}(\widetilde{\rho})$. The concurrence $\mathbb{C} = \mathrm{max} \{0, \frac{2}{P}\left(|d_{tpe}|-\sqrt{p_{00} p_{11}}\, \right)\}$ from a measured $\widetilde{\rho}$
then quantifies a lower bound for entanglement such that $\mathbb{C}=1$ for maximal entanglement and $\mathbb{C}=0$ for a pure quantum state  \cite{Wootters1998,Chou2005,Laurat2007}.
With the approximation $p_{00}\approx1-(p_{01}+p_{01}+p_{11})$, the diagonal terms can be determined experimentally by conditional measurements that distinguish between photons scattered by the $L$ or $R$ samples, e.g., by removing the second beam splitter BS 2.
What concerns the coherence term $d_{tpe}$, it has been shown that this can be approximated as $V(p_{01}+p_{10})/2$ \cite{Chou2005,Laurat2007}, where $V$ is the visibility of the interference fringe at detectors B and C. The latter results from having the remitted single photon interfere with itself on beam splitter BS 2 for different phase shifts and can be experimentally determined.  Typically, an additional Si phase shifter or a vibrating crystal are used to mechanically vary the phase between the two arms in an interferometer \cite{Smirnov1996NE,Hasegawa1999}. In what follows we demonstrate a magnetic, non-mechanical solution for phase modulation that directly and locally controls the nuclear dynamics in each ensemble and can provide an indication of controlling entanglement between the two remote parties.

The rotating orange arrows in Fig.~\ref{fig1}(c-d) depict
the time evolution of the nuclear transition current matrix
elements as defined Ref.~\cite{Shvydko1996}. The nuclear currents in the two crystals are simultaneously driven by  the down-converted  signal photon. Due to the Zeeman shifts $\pm\hbar\Delta_{\bf B}$, the two pairs of nuclear currents in the two samples evolve in directions determined by the sign of the corresponding energy shift and accumulate a phase $\phi(\tau)=\int_0^{\tau}\Delta_{\bf B}(t)dt=\Delta_{\bf B}\tau$ as a constant $\Delta_{\bf B}$ is introduced.
For a pair of currents in a certain crystal, a phase jump of $-2\phi$, associated with a time reversal effect \cite{Shvydko1995}, can be induced by inverting one of the applied magnetic fields at $\tau=\mathrm{T_\phi}$ \cite{Liao2012a,Liao2014}. 
As only $\bf{B_R}$ is inverted at $t=\mathrm{T_\phi}$, the right mode turns into $\cos(\phi-\Delta_{\bf B}t)$ that corresponds  to $\cos(\phi+\Delta_{\bf B}t+\Phi)$ with a phase jump $\Phi=-2\phi$, whereas
the left wavepacket is still proportional to $\cos(\phi+\Delta_{\bf B}t)$. The interference fringe can be analyzed as following \cite{Agarwal2012}
\begin{eqnarray}\label{inout}
\left( 
\begin{array}{c}
 \widehat{a}_{out}    \\
 \widehat{b}_{out}    
\end{array}  
\right) 
&=& 
\frac{1}{2}
\left( 
\begin{array}{cc}
 1  & i   \\
 i  & 1   
\end{array}  
\right)
\left( 
\begin{array}{cc}
 -1  & 0          \\ 
 0   & -1    
\end{array}  
\right) \nonumber \\ 
&\times&
\left( 
\begin{array}{cc}
 \psi_R(t)  & 0          \\
 0          & \psi_L(t)    
\end{array}  
\right) 
\left( 
\begin{array}{cc}
 1  & i   \\
 i  & 1   
\end{array}  
\right)
\left( 
\begin{array}{c}
 \widehat{a}_{in}    \\
 \widehat{b}_{in}    
\end{array}  
\right)\, . 
\end{eqnarray}
Here, $\psi_R(t)=\frac{\psi(\mathrm{T_\phi}+t)}{\cos[\Delta_{\bf B}(\mathrm{T_\phi}+t)]}\cos[\Delta_{\bf B}(\mathrm{T_\phi}-t)]$ and $\psi_L(t)=\psi(\mathrm{T_\phi}+t)$.
Matrices on the right hand side of Eq.~(\ref{inout}) in turn correspond to the action of beam splitter BS 2, mirror, NFS in samples $L$ and $R$  and beam splitter BS 1 on the incident XPDC field. As the field $ \widehat{b}_{in}$ is in the vacuum state, the intensities $Q_B=\int_0^\infty\langle\widehat{a}_{out}^\dagger\widehat{a}_{out}\rangle dt\propto\sin^2\phi$ and $Q_C=\int_0^\infty\langle\widehat{b}_{out}^\dagger\widehat{b}_{out}\rangle dt\propto\cos^2\phi$ at detector B and C, respectively, are plotted in Fig.~\ref{fig1}(e) with $\alpha=1$, $\Gamma=1/141$ GHz for $^{57}$Fe and $\Delta_{\bf B}=30 \Gamma$. Because of the collectively enhanced decay \cite{Roehlsberger2004}, the absorbed x-ray photon is not likely to be retained in the nuclear ensemble as excitation longer than the lifetime of single nuclear excited state. The coherence time of the entanglement between two crystals is approx.~60 ns in Fig.~\ref{fig1}(e).
 However, it has been shown that the speed-up decay can 
be coherently turned off and on via a sequence of rotating \cite{Shvydko1996} or switching off the hyperfine magnetic field \cite{Liao2012a}. While Ref.~\cite{Liao2012a} remains so far only a theoretical proposal, Ref.~\cite{Shvydko1996}
reports the successful  experimental demonstration of  prolonging the nuclear exciton lifetime to $1/\Gamma$ by rotating an externally applied magnetic field of 10 G which in turn controls the internal hyperfine magnetic field inside a $^{57}$FeBO$_3$  crystal. Such a scheme could also be used to extend the coherence time of the presented entanglement setup by using $^{57}$FeBO$_3$ crystals as nuclear samples and a magnetic field rotation setup.

We now proceed with estimates on the possible production rate of heralded macroscopic entanglement. The key requirement here is that the XPDC source produces down-converted x-ray signal photons with energies within the width of the nuclear excited state, where the nuclear resonance absorption exceeds by orders of magnitude  the atomic background processes \cite{Hannon1999}. With a resonance cross section of $\sigma=2.5$~Mbarn for the 14.4 keV transition of $^{57}$Fe, already a nuclear sample of 20~$\mu$m thickness is likely to absorb all incoming resonant photons. Assuming 100 \% detection efficiency \cite{Vagizov2014}, the flux $R_E$ of produced signal photons within the nuclear linewidth equals the rate of heralded entanglement creation. The flux can be estimated as $R_E= \xi_s \Delta E_n/\Delta E_s$, where $\Delta E_n=29.3$ neV is the linewidth of the considered $^{57}$Fe nuclear transition, and 
$\Delta E_s=1$ eV and $\xi_s$ are the bandwidth  and the flux, respectively, of the down-converted signal photons \cite{Tamasaku2007}.
According to Ref.~\cite{Tamasaku2007,Tamasaku2009}, $\xi_s\propto\vert\vec{\chi}^{(2)}_{111}\vert^2  I_p$, where $I_p$ is the photon density of the pump field, and $\vert\vec{\chi}^{(2)}_{111}\vert$ the 111 Fourier coefficient of the second order nonlinear susceptibility for a diamond (111) crystal \cite{Freund1970,Freund1972}. By introducing $\omega_p=\omega_s+\omega_i$ \cite{Tamasaku2007} and the law of cosines \cite{Lennart2004}, we obtain for the susceptibility
\begin{equation}
\vert\vec{\chi}^{(2)}_{111}\vert\approx\frac{Ne^3 F^V_{111}\left(c^2 \vert\hat{Q}_{111}\vert^2-4\omega_s \omega_i\right)}{4 c \varepsilon_0 m^2 \omega_s\omega_i^2 \left(\omega_s^2-\omega_i^2\right)}, 
\end{equation}
where $\omega_p$, $\omega_s$, and $\omega_i$ are the angular frequencies of pump, signal and idler photons, respectively, $N$ is the number density of unit cells, $F^V_{111}$ the linear structure factor of bound electrons \cite{Tamasaku2007,Freund1970} and $\hat{Q}_{111}$ the 111 reciprocal lattice vector of the XPDC diamond crystal. 
Further parameters are $m$ the electron mass, $e$ the electron charge, $c$ the speed of light and $\varepsilon_0$ the vacuum permittivity. Given $\hbar\omega_s=14.4$ keV and $\hbar\omega_i=100$ eV, $\vert\vec{\chi}^{(2)}_{111}\vert\sim 10^{-20}$ C/N $\sim 10^{-16}$ statcoulombs/dynes, having the same order of magnitude as for the case of $\hbar\omega_s=10.9$ keV reported in Ref.~\cite{Tamasaku2007}. Since for the latter SR pulses were used as pump field, the pump photon density can be enhanced by considering an XFEL pulse. Fortunately, diamond crystals are robust and do not experience lattice damage from exposure to intense XFEL radiation \cite{Shvydko2010}. Considering a train of XFEL pulses with $10^{12}$ photons/pulse and repetition rate $f=2.7\times 10^4$ \cite{xfel}, on a spot size of 0.0005 mm$^2$, we obtain $I_p=5.5\times 10^{18}$ photons/s/mm$^2$.  By simple scaling we then obtain $\xi_s=2.9\times 10^{6}$ signal photons/s with a bandwidth of 1 eV, resulting in a production rate 
$R_E$ of around 1 Hz for the heralded creation of entanglement. We note that the signal photon rate is low enough to allow sufficient potential recording time (several hundreds ns) between single shots. Further attention is required for avoiding losses by 
air absorption of the heralding EUV photon \cite{Tamasaku2007} and also for the mechanical alignment of the setup, with XPDC source, beam splitters and mirrors all having angular acceptances of $\mu$rad \cite{Tamasaku2007,Shvydko2010,Hasegawa1995,Hasegawa1999}.

Our scheme for heralded generation of quantum entanglement between two macroscopical nuclear sample crystals relies on M\"ossbauer nuclear transitions. In practice,  quantum effects of the collective nuclear excitation have been shown to be preserved or even induced by vibrating nuclear crystals \cite{Smirnov1996NE,jex1997,Hasegawa1999,Vagizov2014}. Nuclear resonant inelastic coherent scattering \cite{Sturhahn,Seto1995}, for instance, would allow the creation of entanglement in the mechanical motion of a  macroscopic  
system similar to the results reported in Ref.~\cite{Lee2011}, but with increased coherence time and several orders of magnitude increase in the number of involved atoms. The additional feature required from our setup is the detection of phonons in the sample, corresponding with meV energy resolution for the signal photon. We expect that heralded entanglement using x-rays and nuclear transitions can thus open a new research avenue for both applied ideas related to quantum technology as well as more foundational studies of the boundary between the quantum and classical worlds.

\bibliographystyle{apsrev}
\bibliography{NFSBS2A}
\end{document}